# Simplicity Effects in the Experience of Near-Miss

**Jean-Louis Dessalles (dessalles@telecom-paristech.fr)**
Telecom ParisTech, 46 rue Barrault,
F-75013 Paris, France

**Abstract**

Near-miss experiences are one of the main sources of intense emotions. Despite people's consistency when judging near-miss situations and when communicating about them, there is no integrated theoretical account of the phenomenon. In particular, individuals' reaction to near-miss situations is not correctly predicted by rationality-based or probability-based optimization. The present study suggests that emotional intensity in the case of near-miss is in part predicted by Simplicity Theory.

## The Near-Miss Experience

People are emotionally responsive to situations in which some benefit was within reach, but has nevertheless been missed. Near-miss (or near-hit) experiences are one of the main sources of strong emotion. The $1000001^{th}$ customer in a store may feel frustrated when the person in front of her gets his shopping cart reimbursed. Failing an entrance examination by only a few points may spoil entire lives. In June 1995, Tim O'Brien shot himself in the head in despair, after he (mistakenly) believed he had lost out on a £2.7m National Lottery jackpot because he had forgotten to renew his ticket. The missed opportunity is not always as evident as in these examples: individuals sometimes make up an alternative situation to compare their actual fate with, *e.g.* when thinking in retrospect (Teigen, 2005).

In studies about reasoning, feelings generated by near-misses are considered to introduce irrational bias (Wohl & Enzle, 2003; Dillon, Rogers & Tinsley, 2006). If rational agents are supposed to rely on objective probability only, then human sensitivity to near-misses indeed reveals a gross departure from rationality. Though the present study confirms the relative irrelevance of probability, we still suppose that the near-miss experience obeys definite laws.

One of the most remarkable characteristics of near-misses is the human ability to recognize them and to extract all relevant parameters that contribute to making the situation emotional. In particular, though the missed situation has often no objective character (as it did not occur), its closeness to the actual situation is treated as objective by people (Kahneman & Varey, 1990). Since they are easily recognized, near-miss situations populate spontaneous conversations, as individuals systematically urge to share such emotional experiences (Rimé, 2005).

In near-misses, the actual situation is compared to a counterfactual one. The counterfactual alternative may be preferable or worse. The former case is associated with bad luck, while the latter generates feelings of good luck. In the present paper, we deal with bad luck situations exclusively. The symmetrical case can however be easily derived.

In a previous study (Dessalles, 2010), the problem has been explored qualitatively in relation to Simplicity Theory. Participants were proposed short stories and were given the possibility to choose some parameter so as to make emotion maximum. For instance, Lucas had to lace up his shoes at 100m/200m/400m from the station, and then missed the train by five seconds. In this kind of story, a majority of participants choose the shortest distance, as predicted by the theory. The present paper offers two significant improvements. First, we concentrate on bare near-miss (*i.e.* without considering possible mutable causes of the failure, as in the Lucas story). Experiments are thus made simpler and more systematic thanks to a change of modality: we use graphical representations of the missing events and then we vary probability or distance to the target. The second improvement concerns the implementation of the theory, which is done more rigorously.

In the next section, we mention the various factors that are known to influence near-miss emotion intensity, though they have not been included in a coherent theory yet. Then several experiments will be described that explore the role of the complexity (or simplicity) of outcomes. Then we give a short summary of simplicity theory and evaluate its predictions in near-miss situations.

## Determining Factors

A variety of parameters have been found to control emotion intensity in the case of near-miss. The most obvious factor is the difference in 'utility' $\Delta v$ between the actual situation $s_1$ and the counterfactual $s_2$ (Teigen, 2005). Another acknowledged (but ill-defined) factor is the spatial or temporal 'closeness' $D$ to the counterfactual (Kahneman & Varey, 1990; Teigen, 1996; Roese, 1997; Pritchard & Smith, 2004). Teigen (2005) represents these effects through the formula: $L = \Delta v / D$, where $L$ stands for emotion intensity. For some authors, the low probability of the actual event is crucial (Rescher, 1995). Other factors include controllability (Roese, 1997) and the mutability (*i.e.* modifiability) of causes (Kahneman & Miller, 1986; Byrne 2002).

We designed a first test to confirm that the "rational" approach, based exclusively on utility and probability, is unable to predict emotional judgment. Participants were asked to rank emotion in three similar situations where a young man broke his leg while on a one-week skiing holiday. A's accident occurred the first day, on the very first run; in case of B, it occurred the third day, at 14h30; for C, it occurred the last day of the week, on the very last run. We may consider that the probability of each accident was equal. Utility, however, is different: $v(A) < v(B) < v(C)$, since C and, to lesser extent, B, could still enjoy their holiday. If utility was the sole factor, B should be thought to be significantly more disappointed than C. Figure 1 shows average reversed emotion rankings (from 3: most maddening, to 1: least maddening) attributed to A, B and C. We can see that the results contradict the prediction based on pure utility. C's situation is judged as emotional as B's (54% judged it less maddening, but 46% found it more maddening). For nearly half of the participants, the singularity of the temporal location of C's accident (the very last run) as opposed to B's (one run among many) more than compensates for the difference in utility.

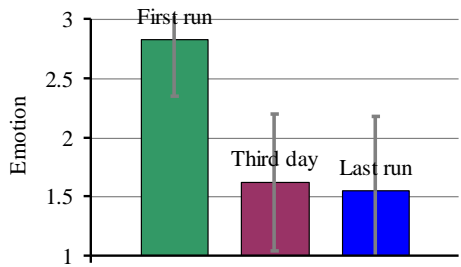

Figure 1: The ski story (102 participants; bars indicate standard deviation).

This result is consistent with Teigen's formula $L = \Delta v / D$, as the 'distance' to the counterfactual (if only C had stopped skiing one run earlier) is minimal for C. Our experiments will show, however, that $\Delta v$ and $D$ are often not the only parameters involved.

We tested various lottery situations in which an individual is supposed to have missed the opportunity of winning 1000 €. The near-miss situations we tested are depicted on figures 2-4. Red areas are winning regions (note that figures 2-e, 3-d and 4-b are not really near-misses). In each case, the "objective" probability of the outcome is evenly distributed over the whole line or plane.

$\Delta v$ is fixed (players were supposed to win 1000 € if the dot had landed in a red region). The way emotion varies is not controlled by the 'objective' winning probability, at least in figures 2 and 4 where that probability is kept constant. It is not entirely controlled by $D$ either, contrary to Teigen's claim, as evidenced by the fact that emotion may significantly vary between figures 2-c and 2-d.

## Experimental Results

Participants (number = 89) were asked to rank the disappointment of losers in various uniform lottery situations. They were shown slides corresponding to figures 2-4 during approximately 1.5 min. In these tests, the dot moved continuously before stopping at its landing site. Each participant was asked to write down different ranks for each test, from 1 (mostly infuriating) to 5 or 6 (least disappointment).

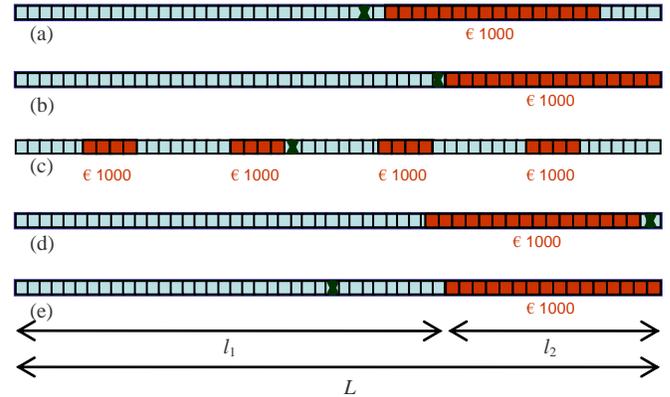

Figure 2: One-dimensional lottery.

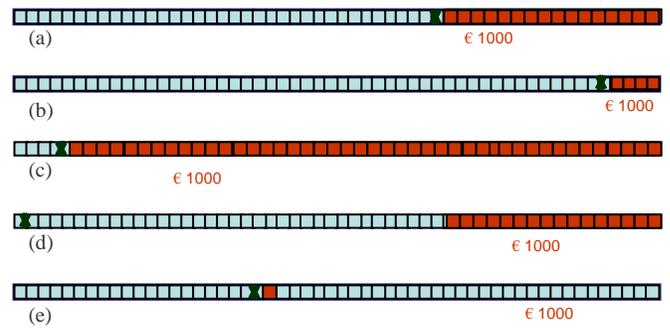

Figure 3: One-dimensional lottery, variable probabilities.

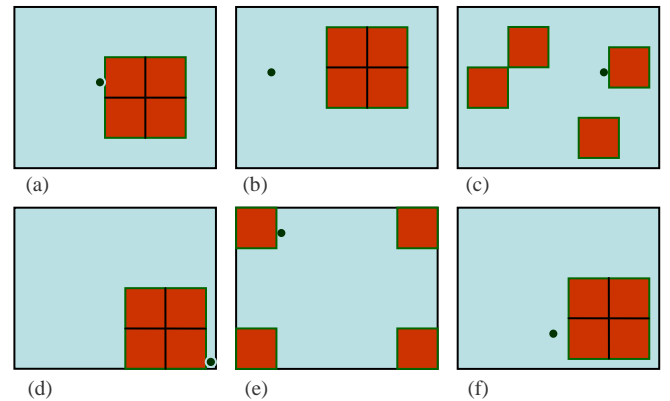

Figure 4: Two-dimensional lottery.

Figures 5, 6, 7 show experimental results for the situations illustrated in figures 2, 3 and 4 respectively. Bars show inverted emotional ranks, ordered from most emotional to least emotional. Segments in grey indicate standard deviation.

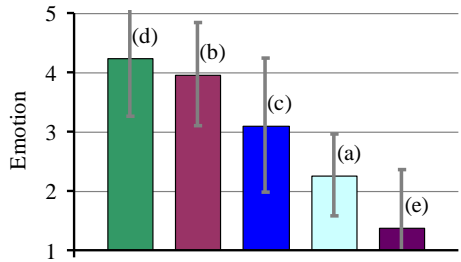

Figure 5: Experiment of figure 2 (89 participants).

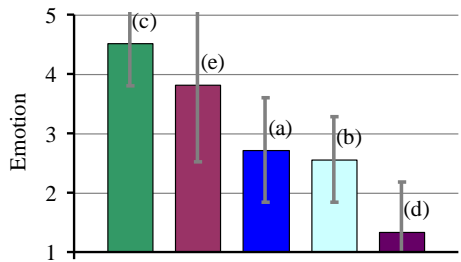

Figure 6: Experiment of figure 3 (89 participants).

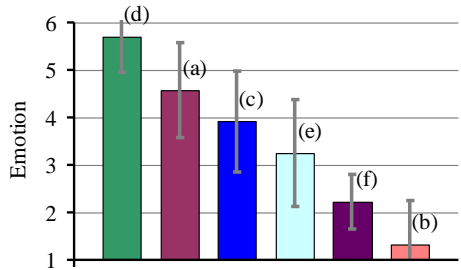

Figure 7: Experiment of figure 4 (89 participants).

These experiments show that probability alone cannot explain emotion ranking, as the winning probability is kept constant in figures 2 and 4, or would make wrong predictions (*e.g.* in figures 3-c and 3-e where emotion is high despite opposite probabilities). More generally, most judgments about (un)lucky situations are not explained by variations of probability (even perceived probability) (Teigen, 1996). The experimental results show that distance to target is not fully relevant either, as judgments for figures 2-c and 4-c show.

We consider now what simplicity theory can possibly bring to the analysis of the problem.

## Simplicity and Probability

Simplicity is a fundamental cognitive principle (Chater, 1999). For instance, it explains how human brains reconstruct hidden shapes (figure 8).

The *description complexity* $C(s)$ of a situation $s$ (or Komogorov complexity) is defined as the size of its (current) best summary. The partially hidden square in figure 8 is simply defined as an invariant of a rotation group, a description that no alternative shape can beat (out of any specific context).

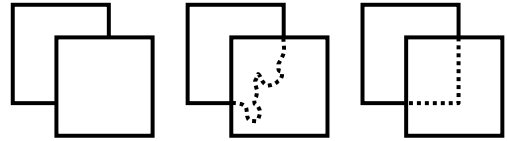

Figure 8: Hidden shapes are the least complex ones.

Complexity theory (Solomonoff, 1978) states that probability is given by:

$$p_a(s) = 2^{-C_w(s)} \qquad (1)$$

$C_w(s)$ is the *generation complexity* of $s$, *i.e.* the minimal amount of information that the 'world' requires to generate $s$. This definition presupposes that $s$ is designated in advance. In most real life situations, this is not the case. Individuals determine probability after the fact. Simplicity theory (Dessalles, 2008) takes the *ex post* determination of $s$ into account by comparing $C_w(s)$ to the complexity $C(s)$ of describing $s$. *Unexpectedness U*, for any situation $s$, is the difference between its generation complexity $C_w$ and the complexity $C$ of its description.

$$U(s) = C_w(s) - C(s) \qquad (2)$$

To be unexpected, situations must be out of the ordinary, i.e. they must be abnormally simple ($C$ smaller than $C_w$). For instance, a lottery draw such as 1-2-3-4-5-6 would be highly unexpected whereas a typical draw would not (what Solomonoff's formula does not predict).

The main claim of Simplicity Theory (ST) is that unexpectedness translates into subjective probability through the following expression.

$$p(s) = 2^{-U(s)} \qquad (3)$$

The difference (in bits) between generation complexity and description complexity, if regarded as successive flips of coin, measures probability. Improbable situations are situations that seem "too" simple, *i.e.* easy to describe and hard to generate. Equations (1) and (3) coincide only when $s$ is fully determined ($C(s) = 0$) before considering its probability (*e.g.* when one's own combination is drawn in a lottery). Note that ST's definition of probability only considers singular events and never sets of alternatives (contrary to standard probability theory).

The notion of unexpectedness explains a variety of phenomena (Dessalles, 2008a; www.simplicitytheory.org). Let's mention a few.

- Rarity: rare situations are felt improbable, but only if the feature that makes them rare is simple enough (a phenomenon that standard probability theory ignores).
- Closeness: events locations, if drawn uniformly, are felt improbable when the location happens to be simple (egocentrically close or close to a simple landmark).

- Exceptions: exceptional situations are considered improbable, but only if the feature that makes them exceptional is simple enough.
- Coincidences: The unexpected character of coincidences is also due to simplicity (Dessalles, 2008b). If $s_1$ and $s_2$ are the two coinciding events, then the complexity of the joint event can be assessed: $C(s_1\&s_2) \leq C(s_1) + C(s_2|s_1)$. Consequently, $U(s_1\&s_2) \geq C_w(s_1) + C_w(s_2) - C(s_1) - C(s_2|s_1)$. If both events are not unexpected separately and have similar generation complexity, we get: $U(s_1\&s_2) \geq C_w(s_1) - C(s_2|s_1)$. Unexpectedness may be large if the analogy between $s_1$ and $s_2$ is strong (which means that the knowledge of $s_1$ allows to spare in the description of $s_2$).

ST's definition of probability can be used to describe aspects of the near-miss subjective experience.

## Simplicity and Near Miss

Let's consider the near-miss situation of figure 9, where the colored zone is the winning one. The actual outcome is $s_1$. It might be compared with a standard (*i.e.* mostly complex) loosing situation $s_{1s}$, or with a standard winning situation $s_{2s}$ or with the closest winning situation $s_2$.

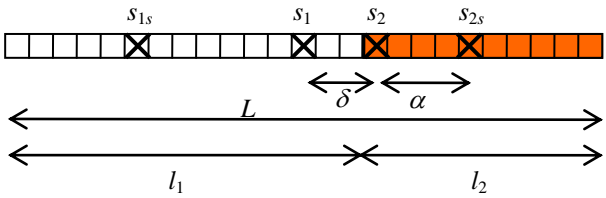

Figure 9: One dimensional near-miss.

The probability of loosing in the lottery is assessed by:

$$U(s_{1s}) = C_w(s_{1s}) - C(s_{1s}) \quad (4)$$

The generation complexity of any landing site $s$ is:

$$C_w(s) = \log(L/a) \quad (5)$$

Logarithms are in base 2 (see appendix for indications about how complexity is computed). This formula reflects the fact that the draw is uniform over the whole strip. On a discrete lottery strip, $a$ is the size of a unitary cell; on a continuous line, it would be the length of the minimal distinguishable landing site. In complexity terms, $C_w(s)$ is the amount of information (in bits) that the 'world machine' needs to produce the event (see www.simplicitytheory.org).

Complexity $C(s)$ is the minimal amount of information needed to designate $s$ unambiguously. Since $s_{1s}$ is a typical position in the losing range, its complexity (see appendix) is: $C(s_{1s}) = \log(l_1/a)$. We get:

$$U_{1s} = U(s_{1s}) = \log(L/l_1) \quad (6)$$

This value is equal, on a logarithmic scale, to the standard ratio of extensional probabilities, though it has been established by considering the complexity of one individual event only. Similarly, we have for a standard winning location:

$$U_{2s} = U(s_{2s}) = \log(L/l_2) \quad (7)$$

Simplicity theory allows different computations for $U(s_1)$. One computation is the straightforward one, given by (6). $U(s_1)$ can be also evaluated by comparison with a counterfactual winning situation $s_2$:

$$C_w(s_2) = C_w(s_1) + C_w(s_2|s_1)$$

$$C(s_2) \leq C(s_1) + C(s_2|s_1)$$

This writing presupposes that $s_2$ is (fictitiously) generated in two steps, through $s_1$. The inequality comes from the fact that any constraint in the computation of complexity may give a suboptimal result. The minimal value of $C(s_2)$ is obtained when $s_2$ is the closest winning position, in which case $C(s_2|s_1) = 0$ ($s_2$ can be determined unambiguously from $s_1$). We get:

$$U(s_1) \geq U(s_2) - C_w(s_2|s_1) \quad (8)$$

$C_w(s_2|s_1)$ measures the smallest amount of information that should be given to the 'world' (here, the lottery machine) for it to produce $s_2$ instead of $s_1$. Normally, $C_w(s_2|s_1) = C_w(s_2)$, as the lottery has no memory. In the case of the near-miss experience, individuals allow themselves to 'cheat' with the world (as the normal world does not allow giving such a bit of a boost), as if they could alter the course of events in retrospect. The value of $C_w(s_2|s_1)$ measures the amplitude of "almost" in the expression "I almost won". It implements the 'distance' $D$ postulated by Teigen (2005). Our definition, however, is more abstract than mere physical distance (sitting next to a lottery winner doesn't necessarily provide a feeling of near-miss!). In the situation depicted in figure 9, we have:

$$C_w(s_2|s_1) = 1 + \log(\delta/a)$$

In that 'cheating' mode, we need 1 bit to 'tell' the world to move the landing site to the right instead of to the left and $\log(\delta/a)$ to designate the amplitude of the move. On the other hand, the complexity of $s_2$ is negligible, as it is a remarkable location. We may consider $C(s_2) = 0$ (see appendix). Finally:

$$U_2 = U(s_1) \geq \log(L/\delta) - 1 \quad (9)$$

We examine now whether the three expressions $Us_1$, $Us_2$, $U_2$ are of any help to account for the specific cases of figures 2-4.

## Simplicity Effects

The main difficulty in applying ST to near-miss situations comes from the fact that we ignore how probability controls emotion. We may assume that emotional intensity is an increasing function of unexpectedness.

A first result is the ability of ST to explain the emotional value of situations depicted in figures 2-d and 4-d. Formula (9) gives a high value for $U(s_1)$, as the missed position is both close to the actual outcome and remarkably simple. In the five situations of figure 2, $a=1$, $L=48$, $l_1=32$, $l_2=16$ and $\alpha \approx l_2/2$. If we apply (9) to situation 2-b, we get

$U_2 = \log (48) - 1 = 4.6$. In situation 2-d, the landing site is blocked at one end of the strip, so we spare the direction bit in the instruction given to the world, and $U_2 = 5.6$. In terms of probabilities, loosing that way is equivalent to hitting a target location designated in advance.

In the two-dimensional situation of figure 4-a, the generation complexity of a draw is $C_w(s) = \log (S/a^2)$, where $S$ is the area of the rectangle. We also have for a standard winning location: $C(s_{2s}) = \log (l_2^2/a^2)$, where $l_2$ is the size of the winning square. For the counterfactual winning position $s_2$, we have: $C(s_2) = \log (4l_2/a)$, as $s_2$ must be located along the perimeter of the winning zone. The complexity of cheating is now: $C_w(s_2|s_1) = 2 + \log (\delta/a)$ (the two additional bits are used to select the 'cheating' direction from four possibilities). Finally:

$$U_2 \geq \log (S/(4l_2)) - 2 - \log \delta \qquad (10)$$

In the experiments of figure 4, $S = 10 \times 8$, $\delta = 0.25$ (or more) and $l_2 = 4$ or 2. We get $U_2 = 2.3$ for situation 4-a and $U_2 \approx 7.3$ for situation 4-d. The difference comes from the fact that $s_2$ is a remarkable point (see appendix) that does not require to be located on the perimeter of the winning zone. Moreover, the world has only two directions available for correcting its 'mistake', instead of four.

Formula (9) also explains the influence of the distance to the target, and thus the systematic preference of 2-b over 2-a over 2-e, and of 4-a over 4-f over 4-b. The formula does more: it accounts for the fact that the counterfactual location $s_2$ is closest to $s_1$, a fact that most models merely take as granted.

Finally, formula (9) explains why situation 2-b dominates 2-c and why situation 4-a dominates 4-c. When the winning region is split, the complexity of the counterfactual $s_2$ increases, by two bits in 2-c and by one bit in 4-c. Unexpectedness is thus diminished in both cases, and emotion is less intense. This is a situation in which subjective probability (as predicted by (3)) changes, while extensional probability does not.

Unfortunately, formula (9) is silent about preferences in figure 3 and it makes two wrong predictions. It wrongly predicts that 2-a should be more emotional than 2-c and that 4-e should be more emotional than 4-c (note that the experimental difference between 4-e and 4-c is not statistically significant). The subjective feeling given informally by some participants is that in 2-c and 4-c, it was "harder" to avoid the winning regions and thus that they are more disappointed. Formulas (6) and (7) may explain these phenomena by providing an estimate of prior probabilities. When the winning region is broken down into four pieces, the complexity of a standard winning region $s_{2s}$ is increased: we need to designate a way of distinguishing the pieces (see appendix). Note that the two additional bits necessary to find the relevant piece are spared when searching into it, as it is four times smaller. But $C(s_{2s})$ gets globally increased and, according to formula (7), winning seems less unexpected and losing appears less probable. Conversely, in a situation like 4-e, the loosing region becomes simpler, making the standard loosing position more complex ($s_{1s}$ can be almost anywhere) and thus making $U(s_{1s})$ smaller than in 4-c. This may explain why the emotion attached to 4-e is relatively downgraded.

## Discussion

This study investigates a phenomenon which, despite its importance in the generation of intense emotions in daily life, resists adequate modeling. In particular, probabilistic models (including Bayesian models) do not provide acceptable explanations.

Simplicity theory does not account for all observed phenomena, but explains for some of them. The principal missing ingredient is the link between probability, as defined by unexpectedness through (3) and emotional intensity. We merely assumed that emotional intensity is an increasing value of unexpectedness. However, we do not know how to integrate prior probabilities, given by (6) and (7), with the amplitude of near-miss given by (9).

The role of prior probabilities is manifest in the situations of figure 3. It is correlated with the fact that extreme values of priors in 3-c and 3-e are felt mostly emotional. However, as evidenced by the large standard deviation in figure 6-e, some participants consider situation 3-e as poorly emotional, as the prior winning probability is very low. We tried to manipulate experimental settings to favor prior probability *vs.* counterfactual thinking. We proposed a version of the experiment in which dots dynamically appeared at random locations before stopping at the near-miss position, instead of moving continuously as in the preceding experiments. The hope was to make people more aware of $U_{1s}$ and $U_{2s}$, as both situations of failures and of success could be observed just before the test began. A comparison of figure 7 and figure 10 shows the consistency of participants' behavior, but fails to show any effect of the prompting.

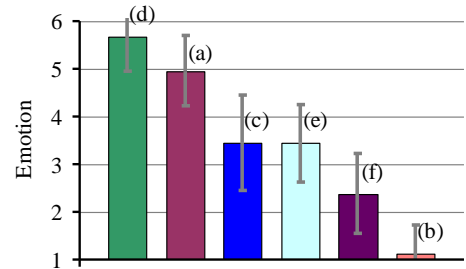

Figure 10: Experiment of figure 4 with random prompting (89 participants).

For some aspects of the near-miss experience, we must perhaps make additional assumptions. In the case of figures 3-c and 3-e, we see that individuals declare equally strong emotions (figure 6) with some indication that 3-c would be even more emotional. Unexpectedness is, however, stronger in 3-e. A possible explanation is that losses provoke more intense emotions than gains (Kahneman & Tversky, 1979).

We are still confident that the near-miss experience obeys definite laws that are in part to be discovered. The aim was to show that simplicity, rather than extensional probability, is the main aspect through which our minds compute emotional intensity. What is at stake is not only the elucidation of an important phenomenon that controls many of our daily emotions. It is also to connect it to a general theoretical framework, Simplicity Theory, which has been developed independently and makes strong predictions in various other domains. One expected outcome of these studies will be to show that emotional judgment is not blurred by a variety of independent biases, but obeys general laws in which simplicity play a central role.

## Acknowledgments

This study is supported by a grant from the "Chaire Modélisation des Imaginaires, Innovation et Création".

## Appendix

Complexity is still regarded as unknowable by many scholars, as it cannot be computed in an "objective" way. This prejudice is not justified in cognitive science (Chater, 1999). Simple codes can approach minimal description. For a list, we can use a positional code.

- 0  1  00  01  10  11  000  001  010 …

Note that the first element requires no additional information (once the list is designated). Such codes are *not* self-delimited. Self-delimitation is irrelevant here, as we are concerned with the minimal description of individual objects. The preceding code can be easily extended to sets of lists (as below) or trees, by using positional code on a branch and switching bits at nodes.

|      | 1     | 11     | 111    |
|------|-------|--------|--------|
| 0    | 10    | 110    | 1110   |
| 00   | 100   | 1100   | 1111   |
| 01   | 101   | 1101   | 11100  |
| 000  | 1000  | 11000  | 11101  |
| 001  | 1001  | 11001  | 11110  |
| 010  | 1010  | 11010  | 11111  |
| 011  | 1011  | 11011  | 111000 |
| 0000 | 10000 | 110000 | 111001 |

The lists may contain structures, operations or even represent short-term memory.

The preceding code can be adapted to assign simple representations to remarkable points. For instance, on a bounded list, endpoints are simpler than middle points.

- 0  00  01  000  001  010  110  101  100  10  1

This explains why remarkable points such as frontiers are less complex than 'normal' points.

With this code, elements of a list of size $N$ are coded with $\log_2 N$ bits on average.